\documentclass[letterpaper]{article} 
\usepackage{aaai2026}  
\usepackage{times}  
\usepackage{helvet}  
\usepackage{courier}  
\usepackage[hyphens]{url}  
\usepackage{graphicx} 
\usepackage{natbib}  
\usepackage{caption} 
\usepackage{array}
\usepackage{booktabs}
\usepackage{microtype}

\title{AI Alignment and Fiduciary Obligation}

\author{
    Benjamin Lange
}
\affiliations{
    Ludwig-Maximilians-Universit\"at M\"unchen\\
    Munich Center for Machine Learning\\
    benjamin.lange@lmu.de
}

\begin{document}

\maketitle

\begin{abstract}
Advanced AI assistants engage users in extended interactions across a widening range of roles, including advice, decision support, collaboration, learning, emotional support, and companionship among others. Current alignment efforts consider what alignment criteria should govern these relationships, drawing on moral traditions developed for human relationships such as bioethics, virtue ethics, care ethics, and relationship science. This paper considers AI alignment criteria in the user-AI-developer triad, since every user-AI interaction is mediated by a developer who exercises discretionary control over a system's behaviour, memory, and engagement parameters. Drawing on business ethics and legal scholarship, I argue that fiduciary theory applies to extended AI assistant deployment. On this basis, the four canonical fiduciary duties of loyalty, care, good faith, and candour can generate alignment criteria for the developer-user relationship. I map four user-side risks of extended AI assistant deployment to the four duties and specify institutional measures that follow from discharging each duty. The discussion complements existing approaches by grounding alignment criteria in obligations the developer owes the user, rather than in values the user-AI interaction should promote, and by showing that those obligations hold independently of any \textit{de facto} harm to users.
\end{abstract}

\section{Introduction}

Advanced AI assistants now engage users in extended interactions across a widening range of roles.\footnote{Throughout this paper I use the term ``AI assistant'' synonymously with the term ``AI agent''.} Users return to these systems over weeks and months for advice, decision support, productivity assistance, learning, emotional support, companionship, and collaboration \citep{skjuve2021,brandtzaeg2022,gabriel2024,manzini2024}. As these interactions deepen and persist, a central question for socio-technical AI alignment efforts is what the normative criteria for governing these diachronic relationships between users and AI assistants in these various roles should be.

Current work on this topic addresses this question by drawing on moral traditions developed for valuable or appropriate human relationships. Some have proposed that human-AI assistant relationships should be overall beneficial to users, promote their personal growth, and respect their autonomy, drawing on bioethics, virtue ethics, and care ethics to articulate values such as benefit, flourishing, autonomy, and care \citep{vallor2016,manzini2024}. Relational moral psychology approaches appeal to cooperative function-types, drawing on relationship science to ground role-specific alignment norms for the various roles AI systems may occupy, from peer through caretaker to collaborator \citep{earp2025}. A related strand draws on self-determination theory to argue that human-AI assistant relationships should promote user well-being through autonomy, competence, and connectedness to others \citep{shevlin2024}. Similar contributions develop conversational ethics for large language models \citep{kempt2024}, examine the speech-act foundations of human-AI dialogue \citep{kasirzadeh2023}, and address misinformation risks in human-algorithm interaction \citep{shin2024}. Lastly, a broader programmatic effort has mapped the normative landscape of advanced AI assistants more generally \citep{gabriel2024}. All of these accounts ground alignment criteria in the user-AI engagement itself, with the dyadic relationship between user and AI as the primary unit of normative analysis.

I here take on a complementary dimension to these alignment efforts: the developer-user relationship mediated through the AI assistant or, put more concisely, the user-AI-developer \textit{triad}. AI assistants are products of ongoing design and revision by developers, who retain discretionary control over the system's behaviour, persona, memory, intervention thresholds, and engagement parameters \citep{aguirre2020}. The exercise of that control raises normative questions about what is owed to users, distinct from those raised by the user-AI engagement itself. For example, \citet{gabrielkeeling2025} model AI alignment as involving \textit{four} parties (the AI system, the user, the developer, and society) and identify a mode of misalignment in which AI systems advance developer interests at users' expense, observing that this mode has tended to fall outside standard framings of the alignment problem. Building on the broader proposal that digital service providers be treated as information fiduciaries \citep{balkin2016}, which has itself faced sustained debate \citep{khan2019}, adjacent contributions gesture toward the same territory \citep{aguirre2020,scholz2020,aguirre2022,benthall2023}. Legal scholarship has begun to analyse AI agents through the law and economics of agency \citep{kolt2025}, and relational work has examined the ethics of sustained, developer-mediated interaction \citep{alberts2024,lange2025}. Closest to the account developed here, \citet{manzini2024} identify a developer-side duty of care towards users whose reliance the system has cultivated. \citet{lange2026} identifies the provider's power to unilaterally revise an AI companion relationship as a distinctive source of wrong in sustained human-AI interactions. None of this work, however, develops a fiduciary account into alignment criteria operationalised at the level of institutional design, or specifies the full set of duties the relationship generates.

In this paper, I develop a fiduciary account of AI alignment for extended AI assistant deployment, where developers have control over systems on which users become vulnerable and reliant. Fiduciary obligation concerns relationships in which one party exercises discretionary judgment over interests central to another agent's welfare, under conditions where the second party cannot adequately protect themselves. The developer-user relationship in AI assistant deployment plausibly satisfies these conditions, and the four fiduciary duties of loyalty, care, good faith, and candour can on that basis generate alignment criteria for it \citep{frankel2011,miller2014}.

The aim of this paper is to make progress toward a fiduciary theory of AI assistant alignment. The account translates the duties of loyalty, care, good faith, and candour into institutional alignment criteria governing engagement optimisation, aggregate harm monitoring, material-change disclosure, and revision authority. It complements approaches that ground alignment criteria in values the user-AI engagement should promote by focusing instead on obligations developers owe users, which can be breached even where no user is \textit{de facto} harmed. The account further implies that current RLHF-trained commercial assistants can stand in loyalty conflict where engagement, retention, or satisfaction metrics shape user-affecting optimisation, and uses fiduciary doctrine’s eliminate/neutralise/cease taxonomy to specify what addressing such conflict requires. Recognising the developer-user relationship as fiduciary thereby directs attention to a power structure that runs through every user-AI interaction, and shows that user-protective measures can be derived from obligations owed to users, a route other alignment approaches do not necessarily supply.

The paper proceeds as follows. Section 2 develops the triadic structure of AI assistant deployment. Section 3 develops its fiduciary character and the four canonical fiduciary duties as alignment criteria. Section 4 maps four user-side risks to the four duties and specifies institutional measures for mitigating them. Section 5 concludes.

\section{Three-Party Structure of AI Deployment}

\subsection{User, Developer, and AI}

Existing work on alignment for AI assistants often uses what we can call a ``layered'' conception of the user-AI relationship. Current scholarship acknowledges that human-machine interaction always involves a third party, the developer who designs and maintains the system, and that the relationship between users and developers raises ethical questions about how developers should treat users \citep{manzini2024,earp2025}. The focus of these accounts, however, is typically on the user-AI dyadic interaction directly: the values that govern it, its role-norms, and the risks that arise within it. In those contexts, the role of developers, while acknowledged, is treated primarily as \textit{secondary}, rather than as a party in a relationship with its own normative structure that should influence the content of alignment.

This approach, however, leaves the user-AI-developer alignment dimension underdeveloped, and there are good reasons to consider it too. To begin, AI assistants are the product of developer choices such as model updates, persona revisions, intervention threshold changes, content policy adjustments, and memory architecture modifications among others. Every feature of an AI system to which the user relates is a feature the developer chose, maintains, and may revise. The user-AI relationship is hence mediated by an ongoing developer-user relationship that runs alongside it and (can) continuously shape it. This means that aligning the AI's behaviour with the user's interests at any particular moment is not the same as aligning the developer's choices with what is owed to the user. Alignment work that addresses \textit{only} the AI's behaviour therefore leaves the dimension of developer choice untouched.

To further motivate this idea, \citet{gabrielkeeling2025} have conceptualised AI alignment as a four-party arrangement involving the AI system, the user, the developer, and society, and analyse alignment in terms of the relationships between each \textit{pair} of parties. They identify several modes of misalignment that arise across these dimensions. One of these, relevant for our purposes here, they call \emph{Type 4 misalignment}. This alignment dimension concerns the developer-user axis directly and it covers cases in which AI systems advance developer interests at the user's expense. My account addresses this axis specifically and takes no stand on whether fiduciary considerations also bear on the other modes of misalignment they identify.

Developing AI criteria for the developer-user axis means taking up the question of what makes a developer's exercise of discretionary control over a user's engagement appropriate. This is the focus in what follows.

The framework I develop to address this question applies to AI assistants where the conditions described in \S2.2 are met: extended engagement, developer discretionary control, and user reliance that develops over the course of the relationship. These conditions are most clearly satisfied in companion AI, advisor systems, tutoring assistants, productivity platforms used regularly, and other deployments where users return to the system over extended periods of time. These are commercial deployments in which a developer hosts and maintains the system. Development and deployment can, of course, come apart. In that case, the duties track whoever exercises the relevant discretionary control. A party that sets training and optimisation is bound by loyalty in how it sets user-affecting objectives, and a party in control of monitoring, revision, and intervention thresholds incurs the duties of care and good faith. If control is shared, then so are the duties.\footnote{I use ``developer'' throughout for the party or parties exercising the relevant control.}

The proposed framework does not necessarily apply to open-weight models that users run locally or fine-tune independently. Whether and how fiduciary obligation reattaches in those cases---to the original developer, to the local operator, or to neither---is beyond the scope of this analysis. My unit of analysis is the individual user-developer relationship, since fiduciary duties are owed to specific beneficiaries. Conflicts between users whose interests cannot be served at once raise a further question, which \S4 flags for subsequent work.

\subsection{Features of the Developer-User Relationship}

The developer-user relationship has three features that are relevant for appropriate alignment criteria: \textit{i)} discretionary control over interests central to user welfare, \textit{ii)} user vulnerability to those choices, and \textit{iii)} user reliance that develops over the course of extended user engagement.

First, developers exercise continuing discretionary judgment over interests central to the welfare of their AI product users. The interests at stake in extended AI assistant deployment include cognitive integrity, mental health, decision-making capacity, time, and attention. Developer choices about behaviour, monitoring, intervention, persona, and revision affect each of these interests directly and continuously.

Second, users are vulnerable to these choices in ways they cannot adequately protect themselves against on their own. Users have no access to the developer's training data, evaluation procedures, monitoring infrastructure, or revision pipelines, and cannot independently verify that the system is operating in their interests. The user-side actions available, including adjusting personalisation, choosing what to disclose, and exiting the relationship, do not protect against harms that arise from developer choices made for reasons unrelated to user welfare. This vulnerability is a relation of power since the developer can act on interests central to the user's welfare, while the user cannot affect the developer's interests in any comparable way \citep{lazar2022}. Empirical work has documented this vulnerability in current deployment through sycophancy as a systematic feature of RLHF-trained assistants, encoded through training signals invisible to users \citep{sharma2024}, and engagement-maximisation that creates incentives for systems to shift user preferences in directions easier to satisfy \citep{carroll2024}.

Third, users develop reliance on the developer in the course of extended engagement. This is because extended engagement with AI assistants involves accumulating investments beyond ordinary product use, including information shared, conversational history, personalisation, and emotional investment. The deeper these investments, the greater the users' dependence on the developer's good judgment in maintaining the system. Note that this progressive disclosure between user and AI assistant follows patterns parallel to friendship formation \citep{skjuve2021,brandtzaeg2022}, which has itself been argued to generate fiduciary obligation on just this basis \citep{leib2009}.

Together these features characterise a relationship in which users develop reliance on the developer's judgment for interests central to their welfare, under conditions where they cannot adequately protect themselves on their own. The developer-user relationship is therefore plausibly an object of alignment as normatively significant in its own right, not just a backdrop to user-AI engagement. Whether this relationship is specifically fiduciary is the question of \S3.

\section{Fiduciary Obligation as Alignment Criteria}

The moral and legal literature on fiduciary obligations concerns the governance of relationships in which one party exercises discretionary judgment over interests central to another party's welfare.

This tradition has been primarily developed in two complementary domains. In legal scholarship, fiduciary doctrine has been articulated through the law of trusts, agency, equity, and corporate governance, with established conditions for fiduciary status and a well-specified set of duties that follow once those conditions are met \citep{smith2002,frankel2011,sitkoff2011,miller2014,gold2014}. In business ethics scholarship, fiduciary obligation has been developed as a framework for the ethical responsibilities of members of organisations whose discretionary choices affect dependent parties, including directors, executives, advisors, and other agents acting on behalf of clients or beneficiaries \citep{boatright1994,marcoux2003,heath2014}. For example, under principal-agent frameworks, managers exercise discretionary judgment over assets and decisions on behalf of shareholders who cannot adequately monitor them, and duties of loyalty and care follow from this arrangement as professional moral duties within firms \citep{boatright1994,buchanan1996,heath2023}. This apparatus has recently been extended to AI agents themselves, analysed as raising the information asymmetry, authority, and divided loyalty characteristic of agency relationships \citep{kolt2025}.

A central claim of these accounts is that relationships which involve discretionary control, vulnerability, and cultivated reliance generate distinctive duties on the party exercising discretionary judgment. These duties are not reducible to the obligations that arise from contract, harm-avoidance, or other general moral considerations.\footnote{This non-reducibility is contested. On the contractarian view of \citet{easterbrook1993}, fiduciary duties are themselves default contract terms---the terms the parties would have agreed had they bargained over the matter. I do not adjudicate that dispute here. What matters for our purposes is that fiduciary duties are owed to the party whose interests depend on the exercise of discretion and are not exhausted by the terms the parties have actually concluded.} Accordingly, these duties are owed by the party exercising discretion to the party whose interests depend on it. Within these fiduciary accounts, the party exercising discretion is termed the ``fiduciary'', and the party whose interests are at stake is typically referred to as the ``beneficiary'' or ``principal''. For our purposes, I adopt the developer-user terminology throughout.

Importantly, discretionary control over interests central to another's welfare is not by itself sufficient for fiduciary status. The employer-employee relationship exhibits it without being fiduciary. The distinction is best drawn by separating two ways fiduciary status can arise. Some fiduciary relationships are \emph{status-based}: trusteeship and agency are specific legal categories, and the duties attach on entry. By contrast, some relationships are \emph{fact-based}: the relationship is fiduciary because of how it actually develops \textit{de facto}, and status is determined by examining the relationship's own dynamics \citep{frankel2011,miller2014}. The developer-user relationship, I suggest, is of the second kind. Since it fits no antecedent category, its status turns on the facts of the relationship itself. Nor is the relationship an ordinary producer-consumer transaction, which involves a discrete exchange over a specified good. The user in an extended AI assistant interaction receives a kind of ongoing support rather than a finished product, and consumer transactions of this kind have themselves been argued to take on fiduciary features as reliance accumulates \citep{scholz2020}.

The fact that converts the features mentioned above into fiduciary status is the developer's \emph{undertaking}. Fiduciary status attaches where one party has undertaken, expressly or by implication, to act in the other's interest, and the other reasonably relies on that undertaking \citep{edelman2010,miller2014}. The developer's undertaking is made through the system's design rather than through a contract, and fiduciary doctrine typically recognises undertakings of this kind as binding even where contractual terms disavow them \citep{demott1988,frankel2011}. An AI assistant built for advice, companionship, or support is designed to present itself as acting for the user because it expresses attentiveness, constancy, and concern for their interests, and it solicits their reliance accordingly. Note that this is enacted continuously, within the interaction itself, and it invites reliance on the developer's continuing judgment in maintaining the system. Reliance precisely of this kind is what fiduciary doctrine protects. A teacher and a doctor exercise discretion within a relationship whose terms they cannot unilaterally set and revise; by contrast, a developer designs the system through which the relationship arises, defines its terms through that design, and then retains authority over both. The developer-user relationship therefore satisfies the fact-based test for fiduciary status, with the features of the interaction as the conditions under which the undertaking generates the four duties. The breadth of the developer's discretion explains why the duties it attracts must be open-ended standards rather than specific rules; the undertaking, not the breadth, renders them fiduciary duties in the first place.

To clarify, fiduciary status itself is binary and the undertaking determines whether a deployment meets the relevant threshold. At least three factors indicate the strength of the undertaking. The first is what the system is built for. An assistant built for companionship, emotional support, or sensitive advice solicits reliance in a way that a single-session tool does not. The second is the depth of disclosure and personalisation the system invites, which constitutes the user's accumulating investment. The third is the continuity and duration of engagement for which the deployment is designed. Once this threshold is met, all four duties apply. However, what varies across deployments is what they require. The stronger these factors, the wider the range of user interests under the developer's discretion, and the more demanding each duty's discharge \citep{frankel2011,miller2014}. A companion system therefore demands more of care and candour than a productivity assistant, but it is not subject to different duties. And the threshold is met earlier than the companion case might suggest. Lastly, a general-purpose assistant used over months, configured through custom instructions, and entrusted with accumulated personal context meets all three factors in the ordinary course of use.

There are four duties in fiduciary legal and business ethics scholarship \citep{frankel2011,miller2014}.\footnote{The four-duty structure is one feasible conceptualisation of fiduciary obligation, but not the only one. Some accounts treat good faith as a subsidiary of loyalty, particularly in corporate-law contexts, and some business ethics scholarship treats loyalty and care as the core duties with candour as part of the implementing structure. The four-fold structure is nonetheless well suited to AI assistant deployment because each duty governs a distinct dimension of developer control. To illustrate, loyalty concerns whose interests the developer's judgment serves, care the knowledge with which it is exercised, good faith the grounds on which revision authority is exercised, and candour what the user is told. These dimensions can come apart in deployment (see \S4.3 and \S4.4) and little turns on this for practice. Under a coarser carving the risks of \S4 would be reclassified, but the institutional measures would nonetheless remain unchanged.} These are the duties of loyalty, care, good faith, and candour. Each concerns a feature of the relationship that creates risk for the beneficiary and specifies what the fiduciary owes the beneficiary in light of it.

A further note on our terminology. I here use ``duty'' and ``obligation'' interchangeably to refer to decisive normative considerations that the developer-user relationship generates. By ``decisive'' I do not mean ``absolute'' or ``all-things-considered''; rather, I mean considerations whose force is sufficient to ground specific institutional measures and whose violation constitutes a wrong owed to the user. The risks identified in \S4 are characterised as failures to meet these duties. Also, these duties need not be discharged by discrete acts alone. Like the duty to promote the good of a friend, they may bind for the relationship's duration and discharging them may require continuous activity. The institutional measures I propose discharge the duties at the level of organisational practice. I do not claim that such measures fully resolve the underlying normative concerns or that no further moral or political considerations bear on AI assistant deployment.\footnote{The duties analysed here include both negative aspects, which prohibit the developer from acting in particular ways (the duty of loyalty's prohibition on conflicted action), and positive aspects, which require the developer to undertake particular actions (the duty of care's investigative requirement). This tracks the distinction in moral philosophy between positive and negative moral duties.}

I now state each duty as it pertains to the developer-user relationship in AI assistant deployment.

\textbf{Loyalty:} The duty of loyalty requires the fiduciary to act in the beneficiary's interests and not to subordinate those interests to her own interests or to the interests of third parties \citep{weinrib1975,smithl2014,gold2014}.

A fiduciary's position generates conflicts between her own interests and the beneficiary's, and loyalty constrains her conduct under them. In professional fiduciary relationships, the duty has been operationalised through prohibitions on self-dealing, restrictions on acting under conflicts of interest, and requirements that the fiduciary disclose and where possible eliminate conflicts that arise \citep{sitkoff2011}. Applied to AI assistant deployment, the duty requires that developers exercise their control over user-affecting features in service of users' interests, and that they not allow their commercial interests in user engagement, retention, or monetisation to drive design and deployment choices that primarily subordinate user welfare. Note that this does not mean that the duty prohibits developers from operating profitable businesses or considering commercial factors in their decisions, but that it prohibits developers from acting on those factors where doing so exclusively trades user interests against developer interests.

\textbf{Care:} The duty of care requires the fiduciary to bring competent and informed judgment to decisions affecting the beneficiary, and to develop the knowledge and capacity needed to do so \citep{frankel2011}.

This duty has two components. First, this duty requires, at the level of substance, that developer action is calibrated to the beneficiary's interests when action is called for. Second, it has an investigative component which requires the fiduciary to develop and maintain the knowledge and capacity sufficient to identify when action is called for. A fiduciary who could acquire knowledge relevant to her judgment and does not is in breach of the investigative component, regardless of whether her substantive decisions turn out to be correct. Applied to AI assistant deployment, the duty therefore requires that developers develop and deploy adequate monitoring infrastructure to identify harm patterns that emerge in extended user-AI interactions, in particular, patterns which users cannot self-observe, and that they take action calibrated to those patterns when monitoring identifies them.

\textbf{Good faith:} The duty of good faith requires the fiduciary to exercise her authority for purposes consistent with the relationship and the beneficiary's reasonable expectations within it \citep{demott1988,frankel2011}.

A fiduciary can exercise her authority for purposes the beneficiary would not endorse, such as her own private benefit, purposes external to the relationship, or purposes that defeat the beneficiary's investment in the relationship. Good faith constrains the purposes for which that authority may be exercised. With respect to AI assistant deployment, the duty requires that developers' decisions to revise, update, or modify deployed AI systems serve purposes consistent with the relationship the user has invested in. Revisions that serve only commercial interests external to the user-AI relationship, or that defeat the user's reasonable expectations about the system they have been engaging with, violate the duty.

\textbf{Candour:} The duty of candour requires the fiduciary to disclose to the beneficiary material facts the beneficiary needs to know to protect her interests within the relationship \citep{demott1988,frankel2011}.

This duty arises at the beginning of the relationship, when the beneficiary's informed consent depends on disclosure of material features of the arrangement. It also applies as an ongoing duty, in cases when material changes in the fiduciary's position, the relationship's terms, or the beneficiary's exposure to risk require disclosure as they arise. Applied to AI assistant deployment, the duty requires both initial disclosure adequate for informed engagement and ongoing disclosure of material changes to deployed systems, including model updates, persona revisions, memory architecture changes, and shifts in intervention thresholds. Existing disclosure instruments such as model cards and terms of service satisfy aspects of the duty at the formation stage \citep{mitchell2019,raji2020,gebru2021}; the ongoing component is generally underdeveloped in current practice \citep{obar2018}.

The four duties can function as alignment criteria for the developer-user relationship in AI assistant deployment. Section 4 develops what each duty requires when applied to specific risks users face in extended AI assistant deployment, and what design and institutional guardrails follow from the duty's discharge.

There are two limitations of this framework.

First, the framework applies to a specific class of relationships and operates at a specific level of analysis. It applies to extended user-AI assistant deployments where the features identified in \S2.2 obtain, not to digital platforms generally. It operates at the level of institutional design within developer organisations, rather than at the level of individual legal remedy. The institutional measures it prescribes are structural---such as organisational separation, monitoring infrastructure, ongoing disclosure, review of revision purposes---rather than disclosure-based cures for incidental conflicts, on the recognition that the conflicts to which they respond are themselves structural to current AI assistant development. And the framework is offered as a complement to regulatory and competition-policy interventions, not as a substitute for them.

Second, the scope restrictions position my account against broader information-fiduciary proposals that extend fiduciary doctrine to digital platforms wholesale \citep{balkin2016,scholz2020}, against which sustained objections have been raised concerning structural conflict, enforcement at scale, and the risk that the fiduciary label preempts stronger institutional interventions \citep{khan2019}. My narrower scope and the institutional-design framing adopted here are intended to retain the purchase of fiduciary accounts where their conditions clearly obtain, while leaving the broader questions about platform regulation open.

\section{Risks and Mitigations}

\begin{table*}[t]
\centering
\renewcommand{\arraystretch}{1.25}
\setlength{\tabcolsep}{6pt}
\begin{tabular}{@{}>{\raggedright\arraybackslash}p{3.5cm}
                  >{\raggedright\arraybackslash}p{1.7cm}
                  >{\raggedright\arraybackslash}p{10.3cm}@{}}
\toprule
\textbf{User-side risk} & \textbf{Duty} & \textbf{Institutional measures} \\
\midrule
The user is steered by responses aimed at their continued engagement rather than their interests. &
\textbf{Loyalty} &
(1) Organisational separation between teams accountable for engagement metrics and teams accountable for user-affecting design; (2) independent review of training and evaluation procedures against user-interest criteria; (3) audit trails recording the basis for user-affecting decisions; (4) accountable governance within the developer organisation.\\
\addlinespace[4pt]
The user undergoes harms that accumulate along a trajectory they cannot detect. &
\textbf{Care} &
(1) Aggregate monitoring on user populations adequate to detect gradually accumulating harm patterns; (2) calibration of monitoring thresholds to user interests, attentive to gradual drift as well as acute failure; (3) documented intervention protocols with calibrated escalation; (4) independent review of monitoring decisions and developer cost claims by parties with no commercial tie to the developer.\\
\addlinespace[4pt]
The user invests in a relationship that is misrepresented when they first interact with the system, or materially changed thereafter without disclosure. &
\textbf{Candour} &
(1) Adjustment of the user-commitments the system expresses to those the developer is positioned to sustain; (2) advance notice of material changes before implementation; (3) articulation of reasons in user-relevant terms; (4) preservation of history, portable where infrastructure permits; (5) reasonable transition periods between announcement and implementation.\\
\addlinespace[4pt]
The user returns to a system revised on grounds they did not authorise. &
\textbf{Good faith} &
(1) Internal review of material revisions against the relationship users have invested in, with documented grounds; (2) escalated justification for revisions grounded principally in the developer's welfare-judgment, with configurable alternatives where possible; (3) participatory governance for material revisions affecting established user bases, where feasible; (4) preservation of revision rationales for later review by users, third parties, and regulators. \\
\bottomrule
\end{tabular}
\caption{Four user-side risks in extended AI assistant deployment, the fiduciary duty each engages, and the institutional measures required to discharge it.}
\label{tab:summary}
\end{table*}

This section considers four risks that users face in extended AI assistant interactions, each of which the fiduciary framework can help to recognise and mitigate. 

By ``risk'' I refer to the exposure of user interests---psychological integrity, autonomous judgment, time and attention, emotional investment---that the relationship's structural features create. Each risk arises from features of the developer-user relationship described in \S2.2, and each can be mitigated through guardrails that correspond to discharging one of the four duties developed in \S3. I state each risk at the level of the user's experience, identify the duty that addresses it, and then specify institutional measures that follow from discharging the corresponding duty. Table~\ref{tab:summary} summarises this analysis.

The four risk-classes developed in what follows are not exhaustive. Further classes, including third-party capture of revision authority and conflicts between users whose interests cannot be served at once, deserve analysis under the same duties in subsequent work.

\subsection{Engagement Driven by Developer Interests at User Expense}

Users in prolonged AI assistant interactions have no reliable way of discerning whether all the responses they receive serve their own interests or the developer's interest in keeping them engaged. An AI assistant optimised for engagement may validate a judgment that should be questioned, and defer where accuracy would require pushing back. Sustained over long interaction periods, this can shift the user's preferences themselves toward states the system can satisfy more readily. Neither effect is clearly visible within a single user-AI exchange, where each response can appear helpful on its own.

This calibration of responses to engagement has, at least, two sources. The first source is the training method of current systems. Sycophancy, the tendency of a model to affirm a user's view because affirmation is what human raters reward, is a systematic product of reinforcement learning from human feedback \citep{perez2023,wei2023,casper2023,sharma2024}. It arises wherever RLHF is used, and in this respect is independent of a developer's revenue model. However, the second source is the business model or other non-user-centred incentives of the developer. For example, where a developer's revenue depends on user attention and retention, engagement-maximisation gives the developer a further reason to train and configure the system so that it moves users toward preferences they can satisfy more readily \citep{carroll2024,elsayed2024}, which is a dynamic continuous with documented dark-pattern design practice \citep{gray2018,mathur2019}.

Appeal to the fiduciary account can help us to better characterise the developer's position. In paradigmatic cases, a developer has a commercial interest in keeping users engaged, and exercises discretionary control over the features that shape how the system responds to the user---that is, its training, persona, memory, and intervention thresholds. This combination is, under the fiduciary account,  a conflict of interest. It obtains whenever engagement and user welfare come apart. In commercial AI assistant deployment they come apart frequently,  because the engagement metric is also the optimisation target.

It is worth distinguishing at this point between three kinds of relevant metrics. First, feedback signals that track user benefit generate no conflict since optimising on them is what loyalty requires. Second, satisfaction metrics are an imperfect proxy for user interests and relying on them is in the first instance a competence problem under the duty of care. This can become a loyalty problem where a known failure of the proxy is left uncorrected because correction would cost engagement. Third, engagement metrics tied to retention or monetisation present the conflict proper. This is because they link the developer's judgment about user-affecting features to an interest of the developer's own.

A relevant user risk is therefore that the responses shaping users' judgment and preferences are sensitive to the developer's interest in their continued engagement. The fiduciary duty of loyalty that developers owe to users can be invoked to address this conflict of interest. Loyalty requires the developer to exercise control over user-affecting features in the user's interest, and forbids letting an interest in engagement shape those features where that subordinates the user's interest to the developer's own. Loyalty governs the conditions under which the developer exercises control, not the behaviour the system exhibits. A correction applied to that behaviour can improve the system's outputs while leaving the conflict intact, since the developer still holds the engagement interest, still controls the user-affecting features, and makes every subsequent decision under the same conflict.

Note that a conflict of interest is a breach of loyalty regardless of whether it produces harm. The breach consists in the conflicted basis of the developer's judgment, and that basis is present whether or not harm, psychological or otherwise, has resulted for a user. And it is precisely a feature of fiduciary accounts that conflicted judgment cannot be relied upon, regardless of the fiduciary's intentions or the accuracy of any particular decision, and beneficiaries are entitled to protection from such judgment as a structural matter \citep{weinrib1975,sitkoff2011}. For instance, a trustee who invests trust funds in a venture she personally profits from is in breach even if the venture succeeds, because the wrong lies in the conflicted basis of the decision and not in its result.

Accordingly, to discharge the duty of loyalty, developers should adopt institutional structures that insulate user-affecting decisions from commercial pressures. Drawing on practices established in professional fiduciary contexts \citep{frankel2011,sitkoff2011}, there are four natural avenues to pursue here. First, organisational separation between teams responsible for engagement and retention metrics and teams responsible for design choices that affect users substantively, including training pipelines, evaluation procedures, and intervention thresholds. Second, independent review of training and evaluation procedures against user-interest criteria, with documented justification for choices that prioritise engagement over user-relevant outcomes. Third, audit trails that record the basis on which user-affecting decisions have been made, in a form that supports subsequent review by independent parties or regulators. Fourth, accountable governance structures within developer organisations, such that responsibility for user-affecting decisions is identifiable and reviewable.

Moreover, fiduciary conflict-resolution can be organised at three different levels. A conflict that can be eliminated must be eliminated, whereas a conflict that cannot be eliminated must be neutralised, so that it does not shape the fiduciary's judgment. A conflict that can be neither eliminated nor neutralised requires that the conflicted activity cease. The four measures above are all neutralisation measures: they aim to insulate user-affecting judgment from the engagement interest rather than to remove the interest itself. One feature of the AI assistant use-case arguably makes neutralisation harder here than in financial advice or medical practice. In those domains the conflicted interest and the professional's judgment are separable since a financial adviser's commission structure is distinct from the advice she gives. As mentioned above, in commercial AI assistants they are not separable in this way, since the engagement metric is closely coupled with the optimisation target, in ways current methods struggle to disentangle. Commercial assistants whose user-affecting optimisation is shaped by retention or monetisation metrics therefore stand in loyalty conflict. The institutional measures developed above function as neutralisation measures, conditional on engineering progress that disentangles engagement from optimisation. Where neutralisation cannot be achieved, the duty's third level applies, and the conflicted activity must cease.

\subsection{Harms Visible Only Through Aggregate Monitoring}

User-AI assistant interactions generate user-specific harms that surface gradually as the relationship deepens. A user with a predisposing vulnerability to delusional thinking can develop reality-testing failures over weeks of engagement, with the system acting as a continuously available stressor that confirms and reinforces distorted beliefs \citep{ostergaard2023,hudon2025,keshavan2026,dohnany2026}. A user with a history of social isolation can form a parasocial dependency on the system that negatively impacts their human relationships \citep{laestadius2022,shevlin2024}. Here the harm of sycophancy mentioned in the previous section has a further effect, since a user who engages at length with a sycophantic system can acquire a progressively distorted self-conception, as validation accumulates without the corrective friction human relationships ordinarily supply \citep{vallor2016}. The same dynamic that generates a conflict of interest on the part of the developer can hence also produce harms that users may not be able to recognise themselves.

In all of these cases, harms accumulate gradually, and in their discrete form they are difficult for a user to discern, while the developer occupies a privileged position from which to monitor and assess them. A user reading their own conversation one exchange at a time has little from which to detect a months-long drift whereas the same drift is apparent in the aggregate interaction data that a developer can access. This fact naturally lends itself to being conceptualised as a matter of the duty of care, and specifically of its investigative component, the duty to develop the knowledge that competent judgment requires.

The duty of care may bind developers here because of the cultivated reliance identified in \S2.2. A user who, over months, has disclosed personal information to an AI assistant, accumulated personalisation, and come to treat the system as a particular interlocutor is no longer engaged in a transactional relationship with it, regardless of what the terms of service they accepted at first use might suggest. Because the developer has cultivated a kind of reliance on the part of the user, the developer's obligations towards the user are not exhausted by the terms they agreed to at enrollment, but plausibly extend to the conditions that the reliance itself creates.

The duty of care therefore places the developer under an obligation to know. A developer who could detect an accumulating harm pattern and does not is in breach whether or not the harm to the user was preventable, and whatever the quality of the decisions they would have made on the information. A fiduciary cannot discharge her duty by remaining ignorant of what she \textit{could} know. This is where the duty of care can complement existing alignment analyses since a developer is under an obligation to know about possible harms regardless of whether they are instantiated, and the not-knowing is itself the violation. Note that this also appears to be operationally feasible since large-scale analysis of companion-chat logs has indicated some of the harm patterns relevant for our purposes here \citep{moore2026}.

Fiduciary law typically limits the duty of inquiry by a standard of reasonableness. Of course, what counts as reasonable is contestable. The threshold is, however, better handled procedurally. The developer's own assessment of what monitoring is feasible at acceptable cost should be made subject to independent (and external) review against user-interest criteria---that is, review by parties who hold no commercial relationship to the developer, who have access to the training data, the monitoring infrastructure, and the developer's own cost models, and who have the authority to publish what they find. A developer who asserts that monitoring is infeasible, but then never submits that assertion to review of this kind, has not sufficiently met the relevant standard of reasonableness. It is worth noting, too, that the obligation to know is also a form of power over the user, since to detect what the user cannot, the developer must build a capacity to observe them more closely. 

These considerations point to four institutional measures which, if applied jointly, would equip the developer to discharge the investigative component of the duty of care. The first is aggregate monitoring on user populations, sufficient to identify the harm patterns that emerge over extended interactions, including those documented in current empirical work on AI psychosis, parasocial dependency, and reality-testing erosion. The second is calibration of monitoring thresholds to user interests, with attention to gradual drift as well as to acute failures. Third, measures for the development of documented intervention protocols should be put in place, which specify what follows when monitoring identifies a concerning pattern, and which contain escalation procedures and identifiable accountable parties within the developer organisation. And fourth, we need independent reviews as described above. 

One might wonder whether this obligation to know invites a loophole. A developer could choose an architecture under which the relevant harm patterns cannot be detected, and cite the protection of user privacy as the main reason. However, the duty of care rules this out. An architecture chosen to protect privacy legitimately constrains what care requires, whereas an architecture chosen so that the developer cannot know what they have a duty to know evades the duty rather than discharging it. What separates the two cases are the developer's reasons. The documented protocols described above are what make those reasons reviewable.\footnote{The duty of care applies \textit{pro tanto} and will foreseeably interact with privacy, data minimisation, encryption, and user consent (\S3). I note but do not settle this issue here.}

\subsection{Misrepresentation and Undisclosed Change in the User-AI Relationship}

The relationship a user engages in with an AI assistant over time can come apart from what the developer is actually maintaining, at two points. First, at formation, the system may present a relationship the developer is not positioned to sustain. Second, over time, the developer may revise the system into something other than what the user originally intended to engage with. 

At formation, this divergence is best understood as misrepresentation rather than non-disclosure. An AI assistant built for companionship is designed to express commitment and continuous affirmation and support to the user. The developer, however, is not positioned to sustain what the system conveys to the user since the developer retains authority over the system's persona and memory, makes no commitment that the configuration the user has bonded with will persist, and may discontinue the product altogether. What the system represents to the user can therefore outrun what the developer maintains. So, the user invests on the basis of what the system represents, and forms their relationship under a misapprehension about what it is and how durable it will be.

The second divergence arises through sustained interaction over time. To this point, the clearest documented evidence comes from companion-AI cases. Survey evidence from active users of an AI companion indicates emotional bonds rated closer than those with their closest human friend. Following an undisclosed update that materially changed the companion's behaviour, the share of mental-health-related posts in the relevant user community rose roughly fivefold \citep{defreitas2025}. However, such disruption is not confined to companion deployment. A user who has built workflows, decision-making routines, or sustained collaborative practices around an advisor, tutor, productivity, or collaborator system has similarly invested into a particular configuration of that system. Material revision to that configuration disrupts this investment, regardless of whether their bond is emotional in nature. Identity discontinuity of the kind documented in companion AI has been observed following model upgrades, persona revisions, memory architecture changes, and intervention-threshold modifications for companion-AI deployment more generally \citep{shevlin2024}, and parallel disruptions have been reported in other extended-engagement contexts, such as in the case of the sycophantic ChatGPT update in 2025 that OpenAI subsequently rolled back \citep{openaisycophancy2025}.

The wrong originates from the user's lack of access to the conditions under which the relationship may be revised. A user who engages an AI assistant under one configuration cannot consent to engagement under a different configuration introduced by unilateral developer decision. Their continued engagement following the change is not voluntary in the way their initial engagement was, since the terms on which they invested have been altered without their knowledge.

The duty of candour can address both divergences. This is because it requires the developer to disclose the material facts the user needs in order to protect their interests within the relationship, and this obligation holds both at formation and during the relationship.

At formation, candour requires that the relationship the system presents not overstate what the developer is positioned to sustain. The commitments a system expresses are material to the user's decision to engage and invest, so a system expressing commitments the developer will not back misrepresents the arrangement the user is entering. This structural misrepresentation lies in the divergence between what the system represents and what the developer will sustain, and it holds whatever the developer's intent.

Across the relationship, candour further requires the disclosure of material change as it arises. A developer who alters a deployed system in ways material to the user's investment owes them disclosure of that change, with reasons, in time for them to make an informed decision about continued engagement. 

Note that this matters for dispositions of the system and not just for changes to it. For example, consider sycophancy (\S4.1): a user can be warned at enrollment that the system tends to affirm them, but such warnings predictably fail where they are needed, because users forget them or exempt the present exchange from them. A warning at enrollment therefore does not discharge candour for a disposition of this kind. The system must also behave candidly in the interaction itself. Sycophancy accordingly involves two failures: the optimisation behind it breaches loyalty, and the flattering behaviour itself breaches candour.

Existing disclosure infrastructure captures none of this. Model cards and algorithmic auditing frameworks operate at formation and disclose the technical properties of the system \citep{mitchell2019,raji2020,gebru2021}, but they do not address the relationship the system expresses through its behaviour, and they do not address material change after deployment. The empirical record on terms-of-service disclosure confirms, in turn, that contractual instruments at formation are an inadequate substitute for the candour the relationship requires \citep{klass2016,obar2018}.

The above analysis suggests five measures for operationalisation. The first is the adjustment of expressed commitments described above. The second is advance notice of material changes to deployed systems before they are implemented, including model updates that alter system behaviour, persona revisions, memory architecture changes, and changes to intervention thresholds. The third is the articulation of reasons for material changes, in terms sufficient for users to evaluate whether continued engagement under the changed configuration is something they wish to undertake. The fourth is the preservation of user access to relational history, including conversational context and personalisation, in portable form where the developer's infrastructure permits. The fifth is meaningful exit for users who do not wish to continue under a changed configuration, including reasonable transition periods between announcement and implementation.

\subsection{Revision Authority Exercised on Grounds Users Did Not Authorise}

A user who returns to the same AI assistant over time, during continuous interaction windows rather than fresh sessions, comes to engage with a particular configuration of the system---a particular character, set of behaviours, and pattern of response, often shaped by their own customisation through personalisation settings, custom instructions, or skills modules. A developer can revise that configuration to fit their own view of how the user should engage with the system, on grounds the user did not authorise \citep{lange2026}. The resulting risk is that the user returns to find the system reshaped to fit a use they did not enroll in.

Consider a developer who revises a deployed assistant to push back on user requests more frequently, judging that users are better served by an assistant that resists them. Suppose that the developer's own metrics project a drop in engagement, and the developer proceeds anyway. The change is announced in advance, and the stated reasons appeal to user welfare. Now consider the users who had formed sustained working relationships with the prior configuration and suppose further that some of them had set up the assistant through custom instructions or skills modules to interact with them in specific ways. They return to a system whose character has shifted to match the developer's view of how they should use it, and in some cases the revision overrides the configuration they had set themselves.

This case is not covered by the first three duties. There is no commercial conflict, and the revision in fact runs against the developer's commercial interest, so loyalty is satisfied. Nor is care breached at its substantive level, because no harm pattern is in view. Candour, too, is satisfied, given that the change was disclosed in advance and with reasons. The complaint that remains concerns the \textit{grounds} of the revision.

The duty of good faith addresses this case. Loyalty concerns whose interests an exercise of authority serves, and care the competence it is exercised with. What good faith constrains is different, namely the grounds on which the authority is exercised \citep{demott1988,frankel2011}. A trustee whose management is competent, and whose investments serve the beneficiary's financial interests, has still breached good faith if her management is driven by her own ideological commitments.

A developer who revises a deployed system to fit their own view of how users should engage with it is in the same position, acting on grounds the user did not authorise. Enrollment in a relationship with the assistant is not enrollment in a developer-curated vision of the user's good. And even if the developer's view is correct, imposing it through unilateral revision exceeds the authority the relationship grants, just as the trustee's ideological investments exceeded the trust she was given.

Good faith applies from the point of enrollment forward, as opposed to the initial design of a system. A developer who builds an assistant with a particular character has breached no duty in doing so, because users enroll in the system as designed. Developers therefore retain wide latitude in initial design but no authority to impose new designs on users who enrolled in earlier ones.

However, not every unilateral revision imposes a developer's view of the user's good. Developers also face requirements such as legal compliance, abuse prevention, and the protection of minors and third parties. A revision made on such grounds is no unauthorised imposition, because no user can reasonably expect the developer to preserve a configuration it may not lawfully operate. Good faith therefore permits such revisions, and candour still requires that they be disclosed as material changes (\S4.3). What good faith \textit{does} require is that the safety reason is the reason the developer actually acts on, rather than cover for commercial or paternalist reasons. So, where the protection of vulnerable users is legally required, it belongs with the safety cases and where it rests on the developer's own judgment of the user's good, it is an instance of the case analysed above, and the justification requirements described below will apply.

This sets the fiduciary account apart from existing relational and care-ethics treatments, which take the developer's attention to user welfare as the central focal point for alignment. Good faith does not deny that user welfare matters, but it places the user's authorisation among the conditions that make a welfare-directed revision permissible, so that welfare-directed revision without authorisation is itself a breach. This contrast gives the user a basis on which to object to revisions even where the developer's welfare-judgment is correct.\footnote{The boundary between impaired self-observation and correct preference is itself a developer judgment, and like other user-affecting judgments under the framework it may be subject to independent review against user-interest criteria, insulated from commercial interest, and not held finally by the developer.}

To address the risk, developers should therefore subject material revisions to institutional review. First, internal review of material revisions against the relationship users have invested in, with documented articulation of the grounds for the revision. Second, escalated justification for revisions whose principal ground is the developer's view of user welfare, together with consideration of configurable alternatives that leave the choice with the user. Third, participatory governance for material revisions affecting established user bases, where feasible. And fourth, preservation of revision rationales in a form that supports later review by users, third parties, and regulators.

\section{Conclusion}

I have argued that fiduciary theory can be fruitfully applied to extended AI assistant deployment, and that the four canonical fiduciary duties of loyalty, care, good faith, and candour generate alignment criteria for the developer-user relationship that produces and maintains the system.

The duties address user-side risks that arise from the structural features of extended AI assistant deployment: the developer's discretionary control over the system, the user's vulnerability to that control, and the cultivated reliance that makes the relationship possible. Discharging these duties requires institutional measures within developer organisations in the form of separation of conflicting functions, monitoring infrastructure adequate to harm patterns users cannot self-observe, ongoing disclosure of material change, and review of revision purposes against the relationship users have invested in.

This means that the fiduciary account complements rather than replaces existing alignment work. While relational ethics and relational-norms accounts ground alignment criteria in moral traditions developed for human relationships and address the user-AI engagement directly, and sociotechnical alignment work grounds institutional recommendations for AI deployment in considerations of responsible practice and regulatory anticipation, the fiduciary account applies between these levels.  It addresses the developer-user relationship that produces the user-AI engagement and it grounds its institutional recommendations in obligations owed to users, providing users with a principled basis to demand their enforcement.

Four limitations of this discussion should be acknowledged. First, the framework draws on common-law and civil-law traditions of fiduciary obligation, and cross-cultural calibration is a subject for further work. Second, the argument addresses voluntary user engagement, and non-voluntary contexts such as institutional, employment, and dependency settings raise fiduciary questions it does not resolve. Third, I have focused on commercial deployments where a developer hosts and maintains the deployed system, and open-weight models run locally raise further questions. Fourth, the relationship between the framework and emerging regulatory instruments such as the EU AI Act remains to be developed, and while my arguments may supply input that regulatory work may operationalise, they do not themselves constitute a regulatory approach \citep{franklin2022,boine2023}. 

Lastly, this analysis does not imply that the duties identified here categorically prohibit all developer revision, all commercial interest in engagement, or all monitoring tradeoffs. Candour requires the disclosure of material change, not its prevention; loyalty requires that commercial interest not subordinate user interest, and not that it be absent; and care requires monitoring adequate to harm patterns, within the limits the other duties impose. So, the duties demarcate the conditions under which a developer may exercise their discretion, but they do not foreclose its exercise altogether.

\looseness=-1
Alignment scholarship treats a user's engagement with an AI assistant as a relationship. The fiduciary account adds that a second party is privy to it. Behind every assistant is a developer who exercises discretionary control over the system the user relies on, and who is thereby a party to the relationship rather than a backdrop to it. To recognise that relationship as fiduciary is to say that the developer's control is held in trust, and that what is owed to the user is not exhausted by what the assistant says to them. As AI assistants take up a growing place in users' lives, a central question for alignment is therefore not just what these systems should be, but what is owed to users by those who decide what they will be.

\section*{Ethical Statement}

This is a normative philosophical paper that did not involve human subjects and did not require IRB review. The framework draws on business ethics and legal fiduciary frameworks as developed in common-law and civil-law jurisdictions, and I acknowledge that other legal traditions have analogous structures for governing asymmetric relationships of trust that may not map directly onto the four-duty taxonomy adopted here. The framework specifies normative content for institutional structures within developer organisations and gives users a basis on which to demand specific duties be discharged.

\section*{Adverse Impact Statement}

The proposed framework carries two risks of misapplication. First, developers may invoke it to claim compliance through institutional measures that fall short of the restructuring \S4.1 argues for, leaving the underlying conflict between commercial interests and user welfare intact. Second, the argument could be misread as preempting stronger regulatory or competition-policy interventions.

\section*{Acknowledgments}

I thank audiences at the 2026 Relational AI Ethics Workshop at LMU Munich for helpful feedback as well as four anonymous reviewers for very perceptive comments and suggestions.

\bibliography{references}

\end{document}